\documentclass[9pt,twocolumn,twoside]{opticajnl}
\journal{opticajournal} 

\setboolean{shortarticle}{false}

\usepackage{lineno}
\usepackage{upgreek}
\usepackage{subcaption}

\title{Integrated Photonic Polarization Synthesizer and Analyzer}

\author[1,*]{Carson G. Valdez}
\author[1]{Anne R. Kroo}
\author[1]{Anna J. Miller}
\author[1]{Charles Roques-Carmes}
\author[1]{David A. B. Miller}
\author[1]{Olav Solgaard}

\affil[1]{Stanford University, Ginzton Laboratory, 348 Via Pueblo Mall, Stanford CA 94305}

\affil[*]{carsongv@stanford.edu}

\begin{abstract}

Polarization-resolved control and measurement of the optical field are essential for a wide range of photonic systems, including coherent communication, polarimetric sensing, and quantum information processing. We present a photonic integrated circuit that enables the generation and analysis of arbitrary polarization states. The device provides reconfigurable access to the full polarization degree of freedom of coherent light within a single integrated platform. We experimentally demonstrate arbitrary polarization state generation spanning the Poincaré sphere, as well as Stokes vector measurement on chip. Unlike conventional Stokes measurements that rely on direct detection, polarization analysis utilizing this architecture is intrinsically non-destructive, preserving the optical signal for further optical domain processing. The devices are fabricated in a commercial foundry using CMOS-compatible processes, enabling scalable and reproducible integration. By combining polarization generation and analysis in a compact and stable photonic circuit, this work eliminates the need for external polarization optics and provides a foundation for robust, polarization-enabled photonic integrated systems.

\end{abstract}

\setboolean{displaycopyright}{false} 

\begin{document}

\maketitle

\LARGE{\textbf{Introduction}}
\normalsize{}

Polarization is a fundamental degree of freedom of light and plays a central role in a wide range of optical systems, from classical and quantum communications \cite{Gisin2002,Savory2010, Kaminow2010} to sensing \cite{Goldstein2011}, imaging \cite{Pusenkova2025, Yang2025, Capasso2019}, and metrology \cite{Falk2014}. The ability to generate, manipulate, and analyze arbitrary polarization states enables advanced functionalities such as polarization-division multiplexing, polarization-resolved coherent detection, ellipsometry, and polarimetric imaging. As photonic systems continue to scale in complexity and performance, precise and reconfigurable control over polarization has become increasingly important.

\begin{figure*}[ht]
\centering
\includegraphics[width=\linewidth]{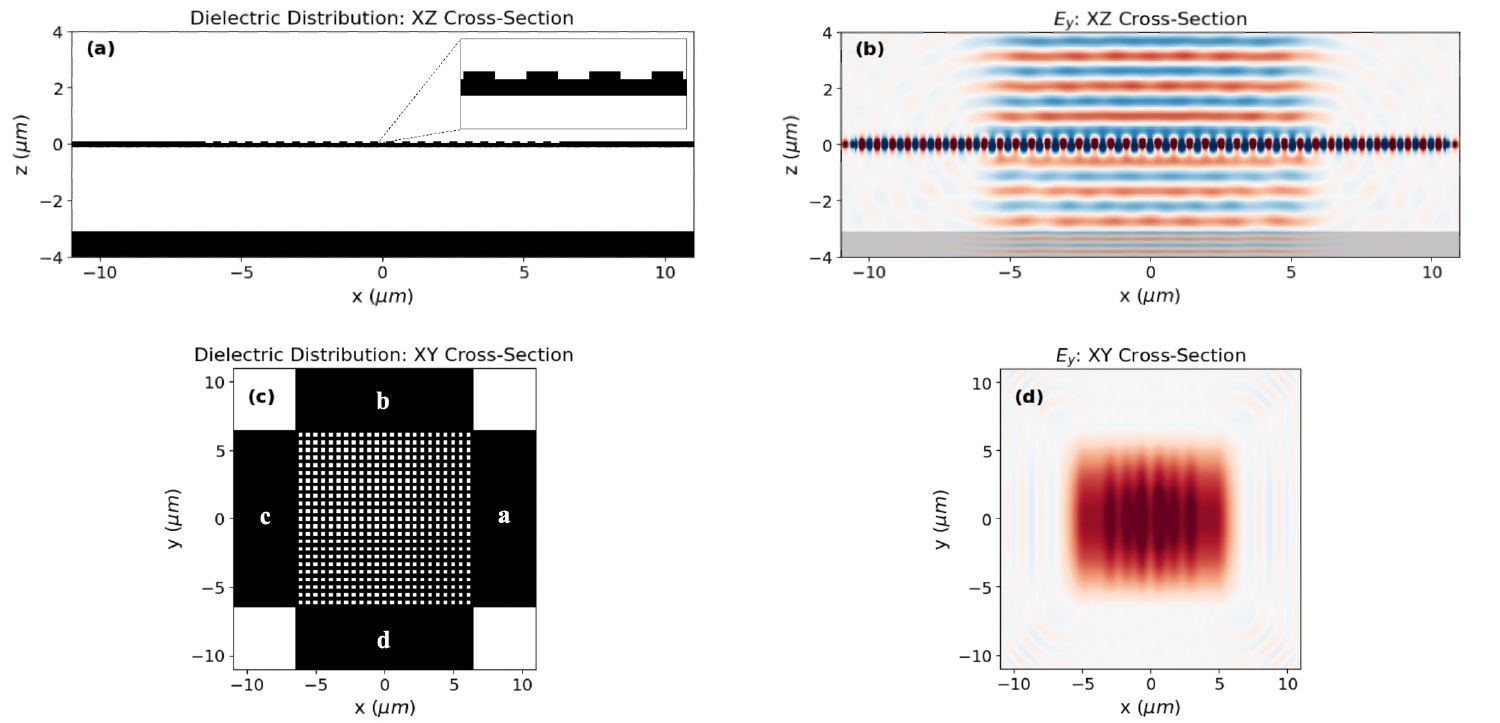}
\caption{\textbf{(a)} X-Z Cross section of the uniform PSGC with a 70 nm partial etch. \textbf{Inset:} Zoom in on a several local periods of the PSGC. \textbf{(b)} X-Z Cross section of the FDTD simulations of the device at 1.550 $\upmu$m with a simulated insertion loss of -2.9 dB. Roughly half of the light is lost toward the substrate as a result of the vertical symmetry of the grating layer. \textbf{(c)} X-Y Cross section of the uniform PSGC showing four-fold symmetry. \textbf{(d)} X-Y Cross section of the FDTD simulations displaying the emission profile of the grating when excited from the left and right ports with equal amplitude, in-phase modes. The emitted grating profile has a 86.4$\%$ mode overlap with a 10.4$\upmu$m MFD gaussian mode.}
\label{fig:psgc}
\end{figure*}


An integrated platform that can both generate and analyze arbitrary polarization states enables a self-contained polarimetric system, eliminating the need for external optics and simplifying system-level integration. Such capability is particularly attractive for applications requiring real-time polarization tracking, adaptive compensation, or compact instrumentation, including coherent transceivers, polarization-sensitive sensors, and on-chip quantum photonic circuits. By bringing polarization control and measurement onto a single chip, polarimetric photonic integrated circuits (PICs) open new opportunities for scalable and robust polarization-enabled photonic systems.

Several methods for integrated photonic polarization analyzers and generators have been investigated previously. In particular, metasurface-based analyzers that operate by spatially demultiplexing polarization components have shown great success \cite{Capasso2019, Zaidi2024, Yang2025, Zhi2019}. However, these devices often act as components in a large bulk optics system. Fully integrated polarization analyzers on PIC platforms have also been demonstrated \cite{Lin19, Lin20, Lin2019_2, Fang21, Banzer2025}; however, these devices operate solely with passive, fixed operation components and lack the reprogrammability required to operate as a polarization generator. Polarization sensitive measurements in beams have also been demonstrated using interferometer meshes fed by sets of grating couplers at different angles. While these systems are capable of performing arbitrary polarization analysis, their spatial resolution is ultimately limited by the array of gratings used as a coupling interface \cite{Butow22}.

We have developed a compact integrated photonic architecture based on reprogrammable, bidirectional photonic meshes for generating and analyzing arbitrary polarization states. The architecture relies on two key components: a polarization splitting grating coupler (PSGC) \cite{Fang21, Luo25, Renyou2023} and a two-stage binary tree photonic mesh comprised of Mach-Zehnder Interferometers (MZIs) \cite{Bogaerts2020, Miller2020, Pai2022, Melloni2024}. The PSGC we have designed is a symmetric, four-port device that performs two operations. It nominally acts as a polarizing beam splitter, spatially demultiplexing the horizontally and vertically polarized components of an input free space beam. Secondarily, the device acts as a polarization rotator, coupling both horizontally and vertically polarized light into the quasi-TE mode of their respective waveguide ports. Thus, the PSGC enables the decomposition of any free space beam into the fundamental quasi-TE mode at each of the device's four ports. The relative complex amplitudes that describe each of the device's four ports then fully characterize the input polarization state.

When operated as an analyzer, we employ a binary tree of MZIs, leveraging self-configuration algorithms \cite{Miller2013, Miller2015, Miller2020} to recombine the outputs of the PSGC into a single output of the PIC. The phase shifter settings required to interferometrically recombine the outputs of the PSGC into a single output port of the PIC contain the information needed to determine the input polarization state. Unlike typical Stokes parameter measurement techniques, this method does not require the direct detection of the beam being analyzed, such that any input signal is preserved for further optical signal processing by spectroscopy \cite{Valdez2025, Miller2025}, spatial mode analysis \cite{Lu2024, Sirbu2025}, or other coherent detection methods \cite{Carmes2024, Rahimi2024}. When operated as a polarization synthesizer, the binary tree of MZIs is used to generate arbitrary sets of complex amplitudes at the ports of the PSGC, resulting in programmatic control of the launched polarization state. In this work, we operate the device in both configurations, demonstrating the efficacy of this architecture for polarization analysis and generation.\newline

\vspace{10pt}
\LARGE{\textbf{Results}}

\large{\textbf{Polarization Splitting Grating Coupler}} 
\normalsize{}

\begin{figure*}[ht]
    \centering
    \includegraphics[width=\textwidth]{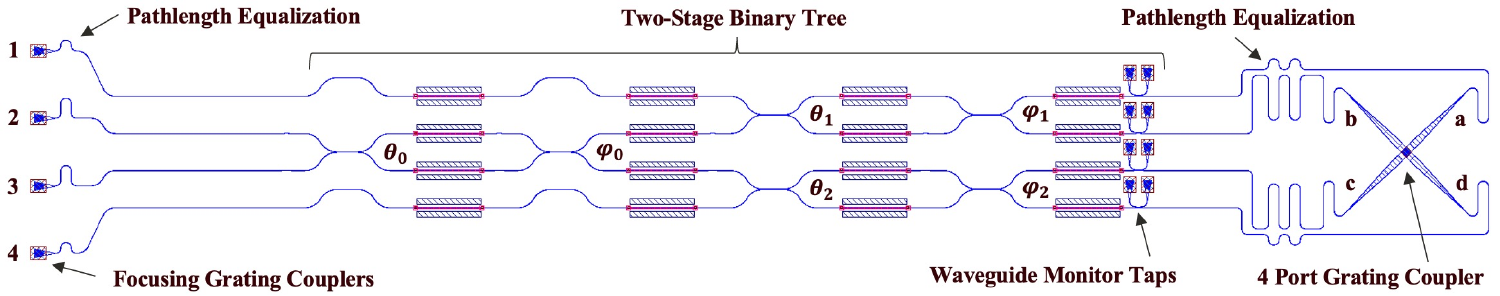}
    \caption{Schematic diagram of the integrated photonic polarization synthesizer/analyzer. The architecture relies on a two-stage binary tree of Mach-Zehnder Interferometers interfaced with the four ports of a normal incidence polarization splitting grating coupler.}
    \label{fig:schematic}
\end{figure*}

The PSGC used in this work is designed by enforcing four-fold symmetry onto a typical uniform Bragg's grating coupler. This device leverages the horizontal symmetry of a normal-incidence beam centered on a two-dimensional square lattice grating coupler. This principle is shown in Fig. \ref{fig:psgc} (a) and (b), where a normal-incidence beam, which has a dominant field component that is out of plane (\textbf{$E_y$}), is centered on a horizontally symmetric grating. It can be observed that the beam is coupled equally into the fundamental quasi-TE mode at the two opposing ports of the device. In accordance with the horizontal symmetry of this system, the two coupled modes will be completely in phase. Likewise, a centered normal-incidence beam that has a dominant field component along the x-axis (\textbf{$E_x$}) will be evenly split, with zero relative phase, between the quasi-TE modes of the opposing vertical ports due to the 90$^\circ$ rotational symmetry of the system.

If the central-symmetric condition of the system is broken by introducing a lateral or angular offset to the beam under analysis, individual ports of the PSGC will receive preferential illumination. In principle, this introduces further degrees of freedom for the analysis and generation of free space beams, but for the purposes of this work, we enforce the symmetric condition. This is achieved in practice by setting the complex amplitude at opposing ports to be equal and in-phase when operating the device as a polarization generator, and by ensuring that any input beam is centered and normally incident when operating the device as a polarization analyzer.

The PSGC is designed for a 220 nm thick silicon on insulator (SOI) platform, with a 3 $\upmu$m thick buried oxide layer. The grating etch depth is set to 70 nm, as determined by the available processes provided by the Multi-Project Wafer (MPW) shuttle run at Advanced Micro Foundry (AMF) Singapore. We employ slowly varying tapers between the PSGC and SOI single-mode waveguides, which reduce the waveguide width from 13 $\upmu$m at the PSGC to 500 nm over a 150 $\upmu$m transition \cite{Fu2014}. The PSGC waveguide width was chosen to be 13 $\upmu$m to closely match the field profile associated with a single-mode fiber at normal incidence. A 50$\%$ fill factor was chosen for the gratings so that both the minimum silicon feature size and the minimum etch feature size would be well within the fabrication tolerances provided by the MPW run. The grating period is determined via the grating equation \cite{Cheng2020}, given in Eq. \ref{eq:refname1}, which imposes phase matching conditions within each unit cell of the grating. Here, $n_{eff}$ represents the effective index of the unperturbed waveguide mode, $n_{pert}$ represents the effective index of the perturbed waveguide mode, $n_{clad}$ represents the index of the cladding, $D$ is the fill factor, $\theta_{grat}$ is the coupling angle of the grating, $\lambda$ is the operating wavelength, and $\Lambda$ is the grating period. Solving Eq. \ref{eq:refname1} given the set of existing parameters at an operating wavelength of 1.55 $\upmu${m} results in a grating period of 556 nm.

\begin{equation}
n_{eff}D + n_{pert}(1-D) + n_{clad}\sin(\theta_{grat}) = \frac{m\lambda}{\Lambda}
\label{eq:refname1}
\end{equation}

The coupler is modeled using a 3-dimensional finite-difference time-domain (FDTD) method with Tidy3D. Figure \ref{fig:psgc} (c) provides a cross section of the modeled geometry in the silicon device layer. The emission profile of the grating is simulated by launching the quasi-TE mode with equal phase from each of the horizontal ports of the PSGC. Figure \ref{fig:psgc} (d) displays a cross section of the simulated field profile 2 $\upmu$m above the device layer, within the oxide top cladding. The emission field profile has an 86.4$\%$ mode overlap with a 10.4 $\upmu$m mode field diameter (MFD) Gaussian field profile, which may be improved in future work by introducing an apodization to the grating design. Furthermore, the coupler exhibits an insertion loss of -2.9 dB at 1.55 $\upmu${m} and a minimum insertion loss of -2.6 dB at 1.535 $\upmu${m}. The efficiency is predominantly limited by large substrate losses, as can be observed in Fig. \ref{fig:psgc} (b), that result from the nearly perfect vertical symmetry of the device, which is only mildly broken by reflections from the silicon substrate and the use of a 70 nm partial etch in the grating layer. In future work, a number of techniques may be employed to break the vertical symmetry and improve directionality at normal incidence, including metallic bottom reflectors, Bragg bottom reflectors, polysilicon overlays, dual-layer grating, and multi-etch blazed grating structures \cite{Cheng2020, Valdez2025_2, Luo25, Renyou2023, Valdez2024, Valdez2023}.


According to the coordinate system we have defined, the complex amplitudes measured through ports \textbf{a} \& \textbf{c}, as shown in Fig \ref{fig:psgc}c, of the PSGC will characterize any horizontally polarized component of the input beam. Similarly, the complex amplitudes measured through the ports \textbf{b} \& \textbf{d},  as shown in Fig \ref{fig:psgc}c, of the PSGC will characterize any vertically polarized component of the input beam. The four ports of the PSGC, represented as a four-element vector of complex amplitudes, fully characterize the polarization of a coupled coherent beam. By analyzing these complex amplitudes, one may determine the polarization state of an input beam; conversely, by setting these complex amplitudes, one may synthesize the polarization state of an output beam. Table \ref{tab:Pol_States} provides the set of complex amplitudes for each port of the PSGC associated with the fundamental polarization basis states.

\begin{table}[htbp]
\small
\centering
\caption{\bf Complex Amplitudes of PSGC Ports}
\begin{tabular}{|| c || c c c c ||}
\hline
Pol. State & Port A  & Port B & Port C & Port D \\
\hline
Horizontal Linear & $\frac{1}{\sqrt{2}}e^{j\theta}$ & 0 & $\frac{1}{\sqrt{2}}e^{j\theta}$ & 0 \\
\hline
Vertical Linear & 0 & $\frac{1}{\sqrt{2}}e^{j\theta}$ & 0 & $\frac{1}{\sqrt{2}}e^{j\theta}$ \\
\hline
+45 Linear &  $\frac{1}{2}e^{j\theta}$ & $\frac{1}{2}e^{j\theta}$ & $\frac{1}{2}e^{j\theta}$ & $\frac{1}{2}e^{j\theta}$ \\
\hline
-45 Linear & $\frac{1}{2}e^{j\theta}$ & $\frac{1}{2}e^{j(\theta-\pi)}$ & $\frac{1}{2}e^{j\theta}$ & $\frac{1}{2}e^{j(\theta-\pi)}$ \\
\hline
Right-Hand Circular & $\frac{1}{2}e^{j\theta}$ & $\frac{1}{2}e^{j(\theta-\frac{\pi}{2})}$ & $\frac{1}{2}e^{j\theta}$ & $\frac{1}{2}e^{j(\theta-\frac{\pi}{2})}$ \\
\hline
Left-Hand Circular & $\frac{1}{2}e^{j\theta}$ & $\frac{1}{2}e^{j(\theta+\frac{\pi}{2})}$ & $\frac{1}{2}e^{j\theta}$ & $\frac{1}{2}e^{j(\theta+\frac{\pi}{2})}$ \\
\hline
\end{tabular}
  \label{tab:Pol_States}
\end{table}


\large{\textbf{Binary Tree Photonic Mesh}}
\normalsize{}

The challenge of measuring and generating complex vectors is well addressed by photonic meshes, which are networks of MZIs arranged in particular architectures \cite{Bogaerts2020, Miller2020}. Within a photonic mesh, each MZI is comprised of a nominal 50:50 beam splitter, an internal tunable phase shifter $\theta$, a nominal 50:50 beam combiner, and an external tunable phase shifter $\phi$. By adjusting the $\theta$ phase shifter of an MZI, one may arbitrarily adjust the splitting ratio of an input to the MZI between its two output ports. By adjusting the $\phi$ phase shifter, one may arbitrarily adjust the relative phase between the two output ports of an MZI. Eq. \ref{eq:refname2} gives the transmission matrix of such an MZI, demonstrating that by adjusting the values of $\theta$ and $\phi$ accordingly, any linear, unitary, 2$\times$2 matrix transformation may be generated. In Eq. \ref{eq:refname2}, $B$ represents the transfer matrix of an ideal 50:50 beam splitter/combiner, $T(\theta)$ and $T(\phi)$ represent the transfer matrices of the internal and external phase shifters, respectively, and $i$ is an indexing variable.

\begin{multline}
T(i) = BT_{\theta}BT_{\phi} \\
 = \frac{1}{\sqrt{2}}\begin{bmatrix} 1 & j \\ j & 1 \end{bmatrix}
\begin{bmatrix} e^{j\theta_i} & 0 \\ 0 & 1 \end{bmatrix}
\frac{1}{\sqrt{2}}\begin{bmatrix} 1 & j \\ j & 1 \end{bmatrix}
\begin{bmatrix} e^{j\phi_i} & 0 \\ 0 & 1 \end{bmatrix} \\
 = je^{j\frac{\theta_i}{2}}\begin{bmatrix} e^{j\phi_i}sin(\frac{\theta_i}{2}) & e^{j\phi_i}cos(\frac{\theta_i}{2}) \\  cos(\frac{\theta_i}{2}) & -sin(\frac{\theta_i}{2}) \end{bmatrix}
\label{eq:refname2}
\end{multline}

Photonic meshes enable the extension of these linear, unitary operators from 2$\times$2 matrix transformations to arbitrarily large N$\times$N transformations \cite{Pai2019}. Here, we employ a two-stage binary tree \cite{Bogaerts2020, Melloni2024, Miller2013}, depicted in Fig. \ref{fig:schematic}, to operate on the four-element vector that represents the ports of the PSGC. The transfer matrix of a binary tree with multiple stages may be represented by the product of matrices that represent each intermediate stage of the mesh. Eq. \ref{eq:refname3} gives the transfer matrix of a two-stage binary tree when operated from left to right, as displayed in Fig. \ref{fig:schematic}.

\begin{multline}
T^{tot} = T^2T^1 \\
= \begin{bmatrix} 
 T_{11}(1) & T_{12}(1) & 0 & 0 \\ 
 T_{21}(1) & T_{22}(1) & 0 & 0 \\ 
 0 & 0 & T_{11}(2) & T_{12}(2) \\
 0 & 0 & T_{21}(2) & T_{22}(2) \\
 \end{bmatrix}
 \begin{bmatrix} 
 1 & 0 & 0 & 0 \\ 
 0 & T_{11}(0) & T_{12}(0) & 0 \\ 
 0 & T_{21}(0) & T_{22}(0) & 0 \\
 0 & 0 & 0 & 1 \\
 \end{bmatrix} \\
= \begin{bmatrix} 
 T_{11}(1) & T_{12}(1)T_{11}(0) & T_{12}(1)T_{12}(0) & 0 \\ 
 T_{21}(1) & T_{22}(1)T_{11}(0) & T_{22}(1)T_{12}(0) & 0 \\ 
0 & T_{11}(2)T_{21}(0) & T_{11}(2)T_{22}(0) & T_{12}(2) \\ 
0 & T_{21}(2)T_{21}(0) & T_{21}(2)T_{22}(0) & T_{22}(2) \\ 
 \end{bmatrix}
\label{eq:refname3}
\end{multline}

From Eq. \ref{eq:refname3} as well as Fig. \ref{fig:schematic}, it can be determined that this architecture of photonic mesh enables an input beam incident on either focusing grating couplers 2 or 3 to be arbitrarily distributed over the ports of the PSGC and vice versa. For the purposes of this work, we will always treat focusing grating coupler 2 as the input/output port of the PIC, although this choice is ultimately arbitrary. Under these conditions, the vector of complex amplitudes that represents the ports of the PSGC can be expressed as functions of $\theta_i$ and $\phi_i$.

\begin{equation}
    \begin{bmatrix} 
 a \\ 
 b \\ 
 c \\
 d \\
 \end{bmatrix} = 
 \begin{bmatrix} 
 -e^{j(\phi_0 + \phi_1 )}e^{\frac{j}{2}(\theta_0 + \theta_1)}\sin(\frac{\theta_0}{2})\cos(\frac{\theta_1}{2}) \\ 
 e^{j\phi_0}e^{\frac{j}{2}(\theta_0 + \theta_1)}\sin(\frac{\theta_0}{2})\sin(\frac{\theta_1}{2}) \\
 -e^{j\phi_2}e^{\frac{j}{2}(\theta_0 + \theta_2)}\cos(\frac{\theta_0}{2})\sin(\frac{\theta_2}{2}) \\
 -e^{\frac{j}{2}(\theta_0 + \theta_2)}\cos(\frac{\theta_0}{2})\cos(\frac{\theta_2}{2}) \\
 \end{bmatrix}
 \label{eq:refname4}
\end{equation}

By controlling the values of $\theta_i, \phi_i$ over a range of $[0,2\pi]$, each element of the complex vector given by Eq. \ref{eq:refname4} may traverse the perimeter of the unit circle in the complex plane. To determine the appropriate values of $\theta_i, \phi_i$, we must consider the constraints of the system. 

When operating as a symmetric coupler with a centered beam at normal incidence, opposing ports of the PSGC will have equal amplitudes and will be in-phase. To satisfy this condition, it is clear from inspection that the first stage of the binary tree should be biased to the quadrature point of the transmission curve such that $\theta_0 = \frac{\pi}{2} \pm 2\pi{n}$. The values of $\theta_1$ and $\theta_2$ are then determined by the ratio of vertically polarized light to horizontally polarized light. In accordance with the coordinate system that we have defined, wherein ports \textbf{a} and \textbf{c} are associated with horizontally polarized light and ports \textbf{b} and \textbf{d} are associated with vertically polarized light,  the phase shifter values $\theta_1$ and $\theta_2$ are given by Eq. \ref{eq:refname5} and \ref{eq:refname6} respectively.

\begin{equation}
 \theta_1 = 2\cos^{-1}(\sqrt{\frac{I_{=}}{I_{||}+I_=}}) \pm 2\pi{n}
 \label{eq:refname5}
\end{equation}

\begin{equation}
 \theta_2 = 2\sin^{-1}(\sqrt{\frac{I_{=}}{I_{||}+I_=}}) \pm 2\pi{n}
 \label{eq:refname6}
\end{equation}

Here $I_=$ denotes the intensity of horizontally polarized light and $I_{||}$ denotes the intensity of vertically polarized light. The fraction of horizontally polarized light can be varied continuously between $[0,1]$ as the values of $\theta_1$ and $\theta_2$ are varied simultaneously between $[0, \pi]$ and $[\pi, 0]$ respectively. 

\begin{figure*}[ht]
\centering
\includegraphics[width=\linewidth]{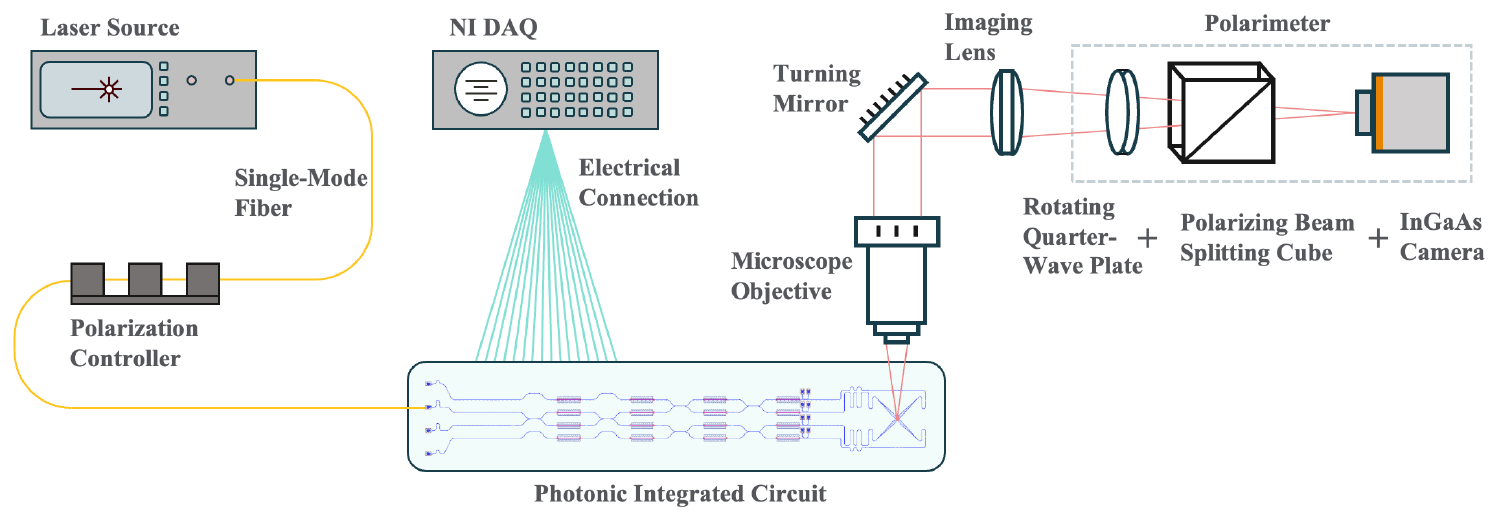}
\caption{Experimental setup for characterizing the PIC as an arbitrary polarization synthesizer. The fundamental quasi-TE mode of the PIC is excited via a fiber coupled external laser source. The phase shifter settings of the photonic mesh are managed via a National Instruments DAQ which allows for complete control of the complex amplitudes entering the four ports of the PSGC. The output of the PSGC is imaged onto a polarimeter which operated on the principles of the rotating quarter-wave plate method.}
\label{fig:sythesizer}
\end{figure*}

In order to maintain constructive interference of the vertically polarized components between ports \textbf{b} and \textbf{d}, the value of $\phi_0$ must be set to compensate for the phase difference introduced by the settings of $\theta_1$ and $\theta_2$. By taking the difference in the angular arguments of components \textbf{b} and \textbf{d} in Eq. \ref{eq:refname4}, the following phase matching condition may be defined.

\begin{equation}
 \phi_0 = \frac{\theta_2}{2} - \frac{\theta_1}{2} \pm (2n\pm1)\pi \label{eq:refname7}
\end{equation}

Similarly, the settings of phase shifters $\phi_1$ and $\phi_2$ must be adjusted to maintain constructive interference of the horizontally polarized components between ports \textbf{a} and \textbf{c}. By taking the difference in the angular arguments of components \textbf{a} and \textbf{c} in Eq. \ref{eq:refname4}, the analogous phase matching condition may be defined.

\begin{equation}
 \phi_1 - \phi_2 = \frac{\theta_2}{2} - \frac{\theta_1}{2} - \phi_0 \label{eq:refname8}
\end{equation}

Under the assumption that the phase matching condition described in Eq. \ref{eq:refname7} is satisfied, this may be simplified further to:

\begin{equation}
 \phi_1 - \phi_2 = \pm (2n\pm1)\pi \label{eq:refname9}
\end{equation}

\begin{figure*}[ht]
\centering
\includegraphics[width=\linewidth]{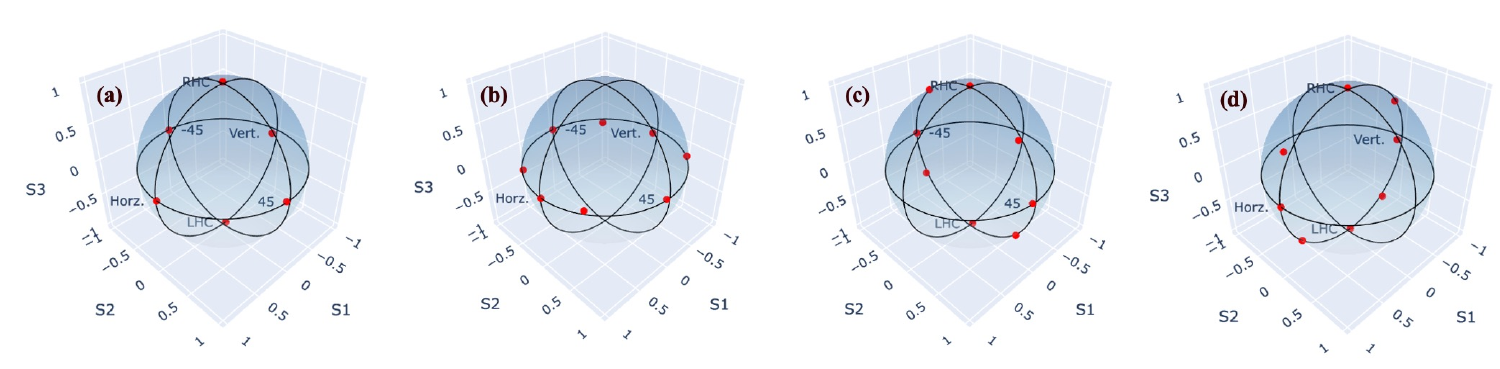}
\caption{Generated states as characterized by the rotating quarter-wave plate polarimeter. (a) Sets of basis pairs demonstrating horizontal and vertical linear polarization states, positive and negative $45^{\circ}$ linear polarization states, as well as right-hand circular and left-hand circular polarization state. (b) Intermediate linear polarization states between the horizontal, vertical, positive and negative $45^{\circ}$ linear poles. (c) Intermediate elliptical polarization states between the right-hand circular, left-hand circular,  positive and negative $45^{\circ}$ linear poles. (d) Intermediate elliptical polarization states between the horizontal, vertical, right- and left-hand circular poles.}
\label{fig:poincare}
\end{figure*}

Finally, from Eq. \ref{eq:refname4}, the relative phase between horizontally and vertically polarized light can be expressed as either $\phi_1$ or $\phi_2$. By adjusting both terms simultaneously, by the same amount and in the same direction, any relative phase between the orthogonal polarization basis may be synthesized. According to these principles, by adjusting the phase shifter settings of the binary tree, any arbitrary polarization state may be mapped between the ports of the PSGC and a single input channel of the PIC. \newline

\large{\textbf{Polarization Synthesizer}}
\normalsize{}

The principles of the arbitrary polarization generator have been tested using the experimental setup detailed in Fig. \ref{fig:sythesizer}. We characterize the output polarization state of the PIC by imaging the emissions of the PSGC onto a rotating quarter-wave plate polarimeter \cite{Wilkinson2021}. For each measurement, the quarter-wave plate is rotated between 0$^\circ$ and 180$^\circ$ in 10$^\circ$ increments while recording the power transmitted through a polarizing beam splitting cube, which acts as a linear polarizer. The relative intensity is recorded using a Bobcat 640 InGaAs camera by integrating over the pixels associated with the output of the PSGC. Each rotation of the quarter-wave plate is repeated over 10 samples for averaging. From the recorded intensity as a function of the quarter-wave plate angle, one can extract the Stokes Parameters for the polarization state under analysis. 

Using this technique, we verify the six fundamental polarization basis states described in Table \ref{tab:Pol_States}. We have plotted the normalized Stokes Parameters on the Poincaré sphere in Fig. \ref{fig:poincare}(a). It can be seen that the device is capable of generating each of the common polarization basis pairs: horizontal and vertical linear polarizations, positive and negative 45$^\circ$ linear polarizations, and left- and right-hand circular polarizations. The accuracy of the polarization generator is broken down into two components. The first is the root mean squared error (RMSE) between the measured portion of horizontally polarized light and the expected value, which we measure as 1.7$\%$ (-17.7 dB). The second is the RMSE of the measured phase difference between horizontally and vertically polarized light, which we estimate to be 0.005$\pi$ radians. It is anticipated that these errors may be reduced in future work through device calibration and error-correction algorithms. 

By appropriately tuning the binary tree mesh, we can generate the superposition of any subset of these basis functions, thus generating any polarization state on the Poincaré sphere. We demonstrate this by synthesizing intermediate polarization states between subsets of each of the basis pairs. In Fig. \ref{fig:poincare}(b), we generate states between the horizontal linear, positive 45$^\circ$ linear, vertical linear, and negative 45$^\circ$ linear polarization states. In Fig. \ref{fig:poincare}(c), we generate states between the positive 45$^\circ$ linear, right-hand circular, negative 45$^\circ$ linear, and left-hand circular polarization states. And in Fig. \ref{fig:poincare}(d), we generate states between the horizontal linear, right-hand circular, vertical linear, and left-hand circular polarization states.\newline

\large{\textbf{Polarization Analyzer}}
\normalsize{}

To demonstrate the polarization analysis capabilities of the PIC, the system is operated in the reverse direction, from right to left as depicted in Fig. \ref{fig:schematic}. Light is coupled into the PSGC via a cleaved single-mode fiber aligned at normal incidence and positioned using a Thorlabs Nanomax 600 series 6-axis stage. Light coupled to the four ports of the PSGC is analyzed using a self-configuration algorithm \cite{Miller2013} based on the power minimization of light out of focusing grating couplers 1, 3, and 4. This algorithm sequentially and iteratively optimizes the phase shifter settings at $\theta_1, \phi_1$ to minimize power at port 1, $\theta_2, \phi_2$ to minimize power at port 4, and $\theta_0, \phi_0$ to minimize power at port 3. The relative intensity at each of these ports is monitored simultaneously by imaging each of the focusing grating couplers onto the InGaAs camera. This process enables us to direct all power from the ports of the PSGC to focusing grating coupler 2 without the need for direct detection of the signal under analysis. The phase shifter settings required to map all power from the four ports of the PSGC to focusing grating coupler 2 provide the necessary information to reconstruct the complex amplitudes at the PSGC via Eq. \ref{eq:refname4}. This eliminates the need for direct detection of the beam output  from focusing grating coupler 2, preserving it for further optical domain signal processing. Further, these algorithms may be continuously run for a form of closed-loop operation, enabling continuous polarization tracking, conversion, and compensation.

\begin{figure}[ht]
\centering
\includegraphics[width=0.7\linewidth]{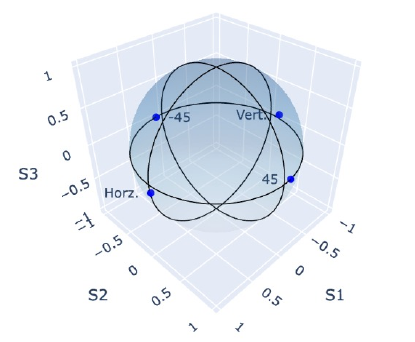}
\caption{Normalized Stokes parameters measured by operating the PIC as a polarization analyzer. The input polarization states are set to horizontal, vertical, positive and negative 45$^\circ$ linear by aligning the input fiber to test structures designed to accept linear polarization states.}
\label{fig:analyzer}
\end{figure}

To launch known polarization states into the PSGC, we use test structures comprising single-polarization grating couplers designed with known rotation angles relative to the PSGC. While the cleaved single-mode fiber is aligned to such a test structure, a manual polarization controller is adjusted to maximize coupling through the single-polarization grating. This enables the accurate alignment of a given polarization state before analysis via the PIC. Figure \ref{fig:analyzer} displays the normalized Stokes parameters extracted via this technique for the horizontal, vertical, positive 45$^\circ$, and negative 45$^\circ$ linear polarizations. The figures of merit for the polarization analyzer are defined similarly to those of the polarization generator. The RMSE for measuring the horizontally polarized component of an input beam and the relative phase between orthogonal polarizations is calculated as 2.9$\%$ (-15.4 dB) and 0.07$\pi$ radians, respectively. Again, it is anticipated that these errors may be reduced in future work by calibrating the devices against a set of known polarization states.

\begin{figure}[ht]
\centering
\includegraphics[width=0.95\linewidth]{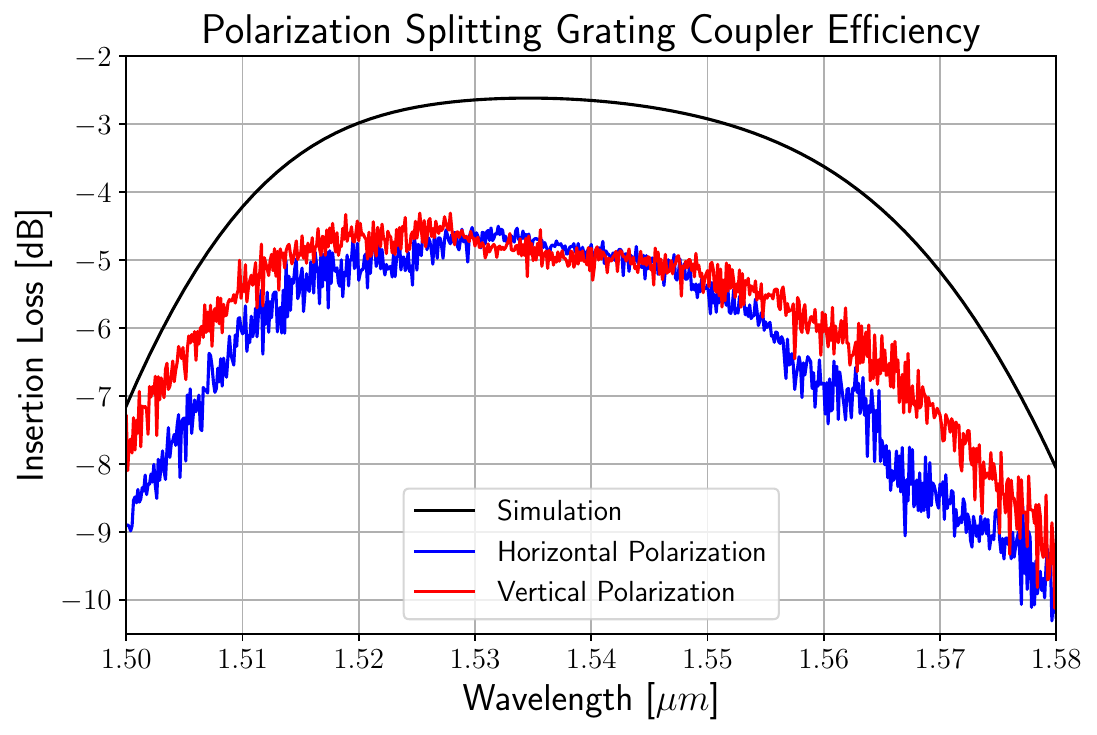}
\caption{Measured coupling efficiency of the normal incidence PSGC when coupled to a single mode fiber for the horizontal and vertical polarization states. We observe minimal polarization dependent loss around the peak of the transmission curve with a minimum insertion loss of -4.5 dB and -4.3 dB for the TE and TM modes respectively. }
\label{fig:efficiency}
\end{figure}

To analyze the coupling efficiency of the PSGC, we once again operate the system in a forward orientation, from left to right as depicted in Fig. \ref{fig:schematic}. Light is coupled into focusing grating coupler 2, while the binary tree is programmed to transmit horizontally polarized light from the PSGC. We record the transmission as a function of wavelength and extract the coupling efficiency by accounting for the input power spectrum of the tunable laser source and the transmission spectrum of the focusing grating couplers. This process is then repeated while the binary tree is reprogrammed to transmit vertically polarized light from the PSGC. The resulting insertion loss measurements are plotted in Fig. \ref{fig:efficiency}. The PSGC exhibits a similar minimum insertion loss for both polarization states, measuring -4.5 dB and -4.3 dB for the TE and TM modes respectively. Additionally, the PSGC maintains a similar 1 dB bandwidth for both polarization states, measuring 42 nm and 45 nm for the TE and TM modes respectively. However, it can be observed that there is a polarization dependent loss (PDL) that increases toward the edges of the operational band, with a maximum PDL of 1.2 dB at 1.565 $\upmu$m. It is possible that the deviation from the expected symmetric coupling of the TE and TM modes is a result of asymmetries introduced in the fabrication process.

\vspace{10pt}
\LARGE{\textbf{Discussion}}
\normalsize{}

We have presented a photonic integrated circuit architecture for arbitrary polarization generation and detection based on a polarization splitting grating coupler, operated at normal incidence, and coupled to a two-stage binary tree of Mach–Zehnder interferometers. Analysis of the circuit reveals the underlying principle by which a reconfigurable photonic mesh can map between a single input port of the PIC and an arbitrary complex vector representing the output ports of the PSGC. This architecture enables flexible and reconfigurable access to the polarization degree of freedom within a compact integrated platform.

When operated as a polarization synthesizer, the phase shifters within the binary tree control both the relative amplitudes and phases of the horizontally and vertically polarized components coupled by the PSGC, allowing the generation of arbitrary polarization states. Conversely, when operated in the reverse direction as a polarization analyzer, the appropriate phase shifter settings are applied to interferometrically route the optical fields from each PSGC port to a single focusing grating coupler on the PIC. In this configuration, the polarization state incident on the PSGC is fully characterized by the phase shifter settings required to achieve constructive interference at the output. Because the polarization information is encoded in these settings rather than in a directly detected optical signal, the architecture naturally enables further downstream optical processing, such as spectroscopy \cite{Valdez2025, Miller2025}, spatial mode analysis \cite{Lu2024, Sirbu2025}, or other coherent detection methods \cite{Carmes2024, Rahimi2024}.

We experimentally demonstrate the performance of the device by synthesizing each of the fundamental polarization basis states, including horizontal and vertical linear polarization, ±45° linear polarization, and left- and right-hand circular polarization. Through characterization of the output polarization states, we determine that the device can accurately control the relative amplitude and phase between horizontal and vertical polarization with RMSEs of 1.7$\%$ (-17.7 dB) and $0.005\pi$ radians, respectively. We further demonstrate the generation of arbitrary polarization states through controlled superposition of these basis states, spanning the full Poincaré sphere. By exploiting the bidirectional operation of the circuit, we also operate the device as a polarization analyzer and experimentally perform Stokes parameter measurements for a set of input polarization states. For the set of polarization states we have analyzed, we estimate the RMSE for the relative amplitude and phase between horizontal and vertical polarization to be 2.9$\%$ (-15.4 dB) and $0.07\pi$ radians, respectively. We expect that in future work, the accuracy of these devices may be improved through calibration efforts, error correction algorithms, and by employing micro-electromechanical (MEMS) phase shifters to eliminate thermal cross-talk between MZIs.

In addition, we characterize the efficiency of the PSGC used to couple between arbitrary free-space polarization states and the quasi-TE mode of the silicon waveguides. We measure minimum insertion losses of -4.3 dB and -4.5 dB for the TE and TM polarization components, respectively. This insertion loss is an improvement over that of the typical focusing grating coupler that we employ, with the added benefit of coupling at normal incidence. In future implementations, the insertion loss may be further reduced through the use of multiple etch steps or additional material layers to break the vertical symmetry. These methods may also serve to improve the accuracy of the device regarding polarization generation and analysis by reducing the impact of secondary reflections and scattering from the substrate.

In conclusion, we have realized a photonic integrated circuit that combines arbitrary polarization generation and polarization analysis within a single integrated platform. By eliminating the need for external polarization optics, this approach reduces system complexity and enables scalable integration with other photonic components fabricated using CMOS-compatible processes. Integrated control and analysis of arbitrary polarization states are particularly relevant for applications that exploit polarization as an information-bearing degree of freedom, including coherent optical communication, polarimetric sensing, and quantum photonic systems. In future work, the scope this of polarization analysis technique may be broadened to address partially coherent light by leveraging recently developed techniques for analyzing spatial and temporal coherency\cite{Miller2025, Carmes2024}. The demonstrated platform provides a foundation for compact, self-calibrated polarimetric photonic circuits capable of real-time polarization tracking and adaptive operation, and highlights the growing role of polarization-enabled architectures in advanced integrated photonic systems.

\vspace{10pt}
\LARGE{\textbf{Materials and Methods}}
\normalsize{}

Our photonic integrated circuits have been fabricated via the commercial provider Advanced Micro Foundry using a multi-project wafer shuttle run. The devices are fabricated on a 220 nm silicon on insulator wafer using the standard available process steps. An HP 81680A tunable laser, set to operate at 1.55 $\upmu$m and coupled to a single-mode fiber, is used as an external source. We use a Thorlabs FPC562 manual polarization controller to adjust the input polarization state. The fiber output of the manual polarization controller has been cleaved and is aligned to the PIC using a Thorlabs Nanomax 600 series 6-axis stage.

The focusing grating couplers used to interface between single-mode fibers and silicon waveguides on the PIC are designed to emit and accept light at close to 12$^\circ$ from normal incidence at 1.55 $\upmu$m \cite{Melloni2024}. These input/output couplers have been characterized to exhibit a minimum insertion loss of -6 dB per coupler at 1.55 $\upmu$m. The silicon single-mode waveguides used to route light throughout the PIC have a 500 nm $\times$ 220 nm cross section, which is maintained throughout the photonic mesh.

Every MZI in the two-stage binary tree employs directional couplers as the nominal 50:50 beam splitter and combiner, each with a 300 nm coupling gap and a 40 $\upmu${m} interaction length. The $\theta$ and $\phi$ phase shifters are implemented as titanium nitride heaters, which rely on the thermo-optic effect to adjust the local refractive index of their respective silicon waveguides. Thermal isolation trenches have been etched through the oxide cladding, silicon device layer, buried oxide, and into the silicon substrate on either side of every thermal phase shifter in an attempt to reduce thermal crosstalk between MZIs. We have placed dummy thermo-optic phase shifters, which are not operational, throughout the mesh to loss match every path through the PIC. Additionally, path-length equalization bends have been introduced throughout the mesh to ensure nominal wavelength independence, propagation loss matching, and coherence retention throughout the PIC.

The phase shifter settings of the binary tree are managed via a National Instruments (NI) Data Acquisition Unit (DAC) with a voltage range between $\pm 10$ V. The $\theta$ phase shifters of the photonic mesh are characterized using a set of monitor taps following the binary tree. Each monitor tap is implemented as a short directional coupler with an approximate $3\%$ coupling strength and a set of focusing grating couplers that redirect the tapped light perpendicular to the chip so that it may be monitored with an overhead infrared camera. To calibrate these $\theta$ phase shifters, the voltage across each heater is sequentially swept while the output of each corresponding monitor tap is recorded. The recorded transmission function is then fitted to the expected intensity functions as determined by Eq. \ref{eq:refname4}. A similar calibration method is implemented for the $\phi$ phase shifters while monitoring the output of the PSGC. Actuation and readout of the calibrated phase shifters enables programmatic generation and analysis of arbitrary polarization states.


\begin{backmatter}
\bmsection{Acknowledgments} This work has been funded by the Air Force Office of Scientific Research (FA9550-21-1-0312, FA9550-23-1-0307), Ames Research Center (80NSSC24M0033), and Stanford Engineering. We thank Flexcompute Inc.for providing access to the Tidy3D software.

\bmsection{Author Details} $^1$Stanford University, Ginzton Laboratory, 348 Via Pueblo Mall, Stanford CA 94305

\bmsection{Author Contributions} D.A.B.M and O.S. supervised the project. D.A.B.M, O.S., C.C., and C.V  developed the analytical model. C.V. designed the PIC. C.V and A.K. built the setups. C.V and A.M performed experimental characterization. C.V wrote the manuscript with input from all authors.

\bmsection{Data availability} Data underlying the results presented in this paper is available from the corresponding authors upon request.

\bmsection{Disclosures} The authors declare no conflicts of interest.

\end{backmatter}

\bibliography{sample}

@article{Capasso2019,
	author = {Noah A. Rubin and Gabriele D'Aversa and Paul Chevalier and Zhujun Shi and Wei Ting Chen and Federico Capasso},
	doi = {10.1126/science.aax1839},
	eprint = {https://www.science.org/doi/pdf/10.1126/science.aax1839},
	journal = {Science},
	number = {6448},
	pages = {eaax1839},
	title = {Matrix Fourier optics enables a compact full-Stokes polarization camera},
	url = {https://www.science.org/doi/abs/10.1126/science.aax1839},
	volume = {365},
	year = {2019}}

@article{Yang2025,
	author = {Yang, Hui and Ou, Kai and Liu, Qiang and Peng, Meiyu and Xie, Zhenwei and Jiang, Yuting and Jia, Honghui and Cheng, Xinbin and Jing, Hui and Hu, Yueqiang and Duan, Huigao},
	date = {2025/01/26},
	doi = {10.1038/s41377-024-01738-1},
	id = {Yang2025},
	isbn = {2047-7538},
	journal = {Light: Science \& Applications},
	number = {1},
	pages = {63},
	title = {Metasurface higher-order poincar{\'e}sphere polarization detection clock},
	url = {https://doi.org/10.1038/s41377-024-01738-1},
	volume = {14},
	year = {2025}}

@article{Zhi2019,
	author = {Jiang, Zhi Hao and Kang, Lei and Yue, Taiwei and Xu, He-Xiu and Yang, Yuanjie and Jin, Zhongwei and Yu, Changyuan and Hong, Wei and Werner, Douglas H. and Qiu, Cheng-Wei},
	doi = {https://doi.org/10.1002/adma.201903983},
	eprint = {https://advanced.onlinelibrary.wiley.com/doi/pdf/10.1002/adma.201903983},
	journal = {Advanced Materials},
	number = {6},
	pages = {1903983},
	title = {A Single Noninterleaved Metasurface for High-Capacity and Flexible Mode Multiplexing of Higher-Order Poincar{\'e} Sphere Beams},
	url = {https://advanced.onlinelibrary.wiley.com/doi/abs/10.1002/adma.201903983},
	volume = {32},
	year = {2020}}

@article{Lin19,
	author = {Zhongjin Lin and Leslie Rusch and Yuxuan Chen and Wei Shi},
	doi = {10.1364/OE.27.004867},
	journal = {Opt. Express},
	month = {Feb},
	number = {4},
	pages = {4867--4877},
	publisher = {Optica Publishing Group},
	title = {Chip-scale, full-Stokes polarimeter},
	url = {https://opg.optica.org/oe/abstract.cfm?URI=oe-27-4-4867},
	volume = {27},
	year = {2019}}

@article{Lin20,
	author = {Zhongjin Lin and Tigran Dadalyan and Simon B\'{e}langer-de Villers and Tigran Galstian and Wei Shi},
	doi = {10.1364/PRJ.385008},
	journal = {Photon. Res.},
	month = {Jun},
	number = {6},
	pages = {864--874},
	publisher = {Optica Publishing Group},
	title = {Chip-scale full-Stokes spectropolarimeter in silicon photonic circuits},
	url = {https://opg.optica.org/prj/abstract.cfm?URI=prj-8-6-864},
	volume = {8},
	year = {2020}}

@article{Lin2019_2,
	author = {Lin, Zhongjin and Rusch, Leslie A. and Chen, Yuxuan and Shi, Wei},
	doi = {10.1063/1.5098492},
	eprint = {https://pubs.aip.org/aip/app/article-pdf/doi/10.1063/1.5098492/13679152/100806_1_online.pdf},
	issn = {2378-0967},
	journal = {APL Photonics},
	month = {10},
	number = {10},
	pages = {100806},
	title = {Optimal ultra-miniature polarimeters in silicon photonic integrated circuits},
	url = {https://doi.org/10.1063/1.5098492},
	volume = {4},
	year = {2019}}

@article{Fang21,
	author = {Liang Fang and Shuang Zheng and Jian Wang},
	doi = {10.1364/OE.437410},
	journal = {Opt. Express},
	month = {Sep},
	number = {20},
	pages = {31026--31035},
	publisher = {Optica Publishing Group},
	title = {Design of on-chip polarimetry with Stokes-determined silicon photonic circuits},
	url = {https://opg.optica.org/oe/abstract.cfm?URI=oe-29-20-31026},
	volume = {29},
	year = {2021}}

@article{Luo25,
	author = {Yannong Luo and Meiyan Wu and Bigeng Chen and Shaoliang Yu and Ping Chen and Shengqian Gao and Renyou Ge},
	journal = {J. Lightwave Technol.},
	month = {Aug},
	number = {16},
	pages = {7777--7783},
	publisher = {Optica Publishing Group},
	title = {Silicon Perfectly-Vertical Polarization Diversity Grating Coupler With Near-Zero Polarization Dependent Loss},
	url = {https://opg.optica.org/jlt/abstract.cfm?URI=jlt-43-16-7777},
	volume = {43},
	year = {2025}}

@INPROCEEDINGS{Renyou2023,
  author={Ge, Renyou and Gao, Shengqian and Wu, Meiyan and Chen, Ping and Chen, Bigeng and Luo, Yannong},
  booktitle={2023 Asia Communications and Photonics Conference/2023 International Photonics and Optoelectronics Meetings (ACP/POEM)}, 
  title={Inverse-Designed Two-Dimensional Grating Coupler with Low Polarization-Dependent Loss}, 
  year={2023},
  volume={},
  number={},
  pages={1-4},
  doi={10.1109/ACP/POEM59049.2023.10368923}}

@article{Miller2020,
	author = {David A. B. Miller},
	doi = {10.1364/OPTICA.391592},
	journal = {Optica},
	month = {Jul},
	number = {7},
	pages = {794--801},
	publisher = {Optica Publishing Group},
	title = {Analyzing and generating multimode optical fields using self-configuring networks},
	url = {https://opg.optica.org/optica/abstract.cfm?URI=optica-7-7-794},
	volume = {7},
	year = {2020}}

@article{Pai2022,
	author = {Sunil Pai and Olav Solgaard and Shanhui Fan and David A. B. Miller},
	journal = {ArXiv},
	title = {Scalable and self-correcting photonic computation using balanced photonic binary tree cascades},
	url = {https://api.semanticscholar.org/CorpusID:253237317},
	volume = {abs/2210.16935},
	year = {2022}}

@article{Bogaerts2020,
	author = {Bogaerts, Wim and P{\'e}rez, Daniel and Capmany, Jos{\'e} and Miller, David A. B. and Poon, Joyce and Englund, Dirk and Morichetti, Francesco and Melloni, Andrea},
	date = {2020/10/01},
	doi = {10.1038/s41586-020-2764-0},
	id = {Bogaerts2020},
	isbn = {1476-4687},
	journal = {Nature},
	number = {7828},
	pages = {207--216},
	title = {Programmable photonic circuits},
	url = {https://doi.org/10.1038/s41586-020-2764-0},
	volume = {586},
	year = {2020}}

@article{Miller2013,
	author = {David A. B. Miller},
	doi = {10.1364/OE.21.006360},
	journal = {Opt. Express},
	month = {Mar},
	number = {5},
	pages = {6360--6370},
	publisher = {Optica Publishing Group},
	title = {Self-aligning universal beam coupler},
	url = {https://opg.optica.org/oe/abstract.cfm?URI=oe-21-5-6360},
	volume = {21},
	year = {2013}}

@article{Miller2015,
	author = {David A. B. Miller},
	doi = {10.1364/OPTICA.2.000747},
	journal = {Optica},
	month = {Aug},
	number = {8},
	pages = {747--750},
	publisher = {Optica Publishing Group},
	title = {Perfect optics with imperfect components},
	url = {https://opg.optica.org/optica/abstract.cfm?URI=optica-2-8-747},
	volume = {2},
	year = {2015}}

@misc{Valdez2025,
      title={Programmable Optical Filters Based on Feed-Forward Photonic Meshes}, 
      author={Carson G. Valdez and Anne R. Kroo and Marek Vlk and Charles Roques-Carmes and Shanhui Fan and David A. B. Miller and Olav Solgaard},
      year={2025},
      eprint={2509.12059},
      archivePrefix={arXiv},
      primaryClass={physics.optics},
      url={https://arxiv.org/abs/2509.12059}, 
}

@article{Miller2025,
	author = {David A. B. Miller and Charles Roques-Carmes and Carson G. Valdez and Anne R. Kroo and Marek Vlk and Shanhui Fan and Olav Solgaard},
	doi = {10.1364/OPTICA.557630},
	journal = {Optica},
	month = {Sep},
	number = {9},
	pages = {1417--1426},
	publisher = {Optica Publishing Group},
	title = {Universal programmable and self-configuring optical filter},
	url = {https://opg.optica.org/optica/abstract.cfm?URI=optica-12-9-1417},
	volume = {12},
	year = {2025}}

@article{Lu2024,
	author = {Lu, Kaihang and Chen, Zengqi and Chen, Hao and Zhou, Wu and Zhang, Zunyue and Tsang, Hon Ki and Tong, Yeyu},
	date = {2024/04/25},
	doi = {10.1038/s41467-024-47907-z},
	id = {Lu2024},
	isbn = {2041-1723},
	journal = {Nature Communications},
	number = {1},
	pages = {3515},
	title = {Empowering high-dimensional optical fiber communications with integrated photonic processors},
	url = {https://doi.org/10.1038/s41467-024-47907-z},
	volume = {15},
	year = {2024}}

@inproceedings{Sirbu2025,
	author = {Dan Sirbu and Ruslan Belikov and Eduardo Bendek and Kevin Fogarty and Rachel Morgan and Kevin Sims and Carson Valdez and Anne Kroo and Marek Vlk and Olav Solgaard and David A. B. Miller},
	booktitle = {UV/Optical/IR Space Telescopes and Instruments: Innovative Technologies and Concepts XII},
	doi = {10.1117/12.3065301},
	editor = {Jonathan W. Arenberg and H. Philip Stahl},
	organization = {International Society for Optics and Photonics},
	pages = {1362309},
	publisher = {SPIE},
	title = {{AstroPIC II: overview of technology development for a near-infrared photonic integrated coronagraph for the Habitable Worlds Observatory}},
	url = {https://doi.org/10.1117/12.3065301},
	volume = {13623},
	year = {2025}}

@article{Carmes2024,
	author = {Roques-Carmes, Charles and Fan, Shanhui and Miller, David A. B.},
	date = {2024/09/20},
	doi = {10.1038/s41377-024-01622-y},
	id = {Roques-Carmes2024},
	isbn = {2047-7538},
	journal = {Light: Science \& Applications},
	number = {1},
	pages = {260},
	title = {Measuring, processing, and generating partially coherent light with self-configuring optics},
	url = {https://doi.org/10.1038/s41377-024-01622-y},
	volume = {13},
	year = {2024}}

@article{Rahimi2024,
	author = {Sadra Rahimi Kari and Nicholas A. Nobile and Dominique Pantin and Vivswan Shah and Nathan Youngblood},
	doi = {10.1364/OPTICA.507525},
	journal = {Optica},
	month = {Apr},
	number = {4},
	pages = {542--551},
	publisher = {Optica Publishing Group},
	title = {Realization of an integrated coherent photonic platform for scalable matrix operations},
	url = {https://opg.optica.org/optica/abstract.cfm?URI=optica-11-4-542},
	volume = {11},
	year = {2024}}

@article{Cheng2020,
	article-number = {666},
	author = {Cheng, Lirong and Mao, Simei and Li, Zhi and Han, Yaqi and Fu, H. Y.},
	doi = {10.3390/mi11070666},
	issn = {2072-666X},
	journal = {Micromachines},
	number = {7},
	pubmedid = {32650573},
	title = {Grating Couplers on Silicon Photonics: Design Principles, Emerging Trends and Practical Issues},
	url = {https://www.mdpi.com/2072-666X/11/7/666},
	volume = {11},
	year = {2020}}

@misc{Valdez2025_2,
      title={Three-Wave Interaction Grating Coupler with Sub-Decibel Insertion Loss at Normal Incidence}, 
      author={Carson G. Valdez and Simon A. Bongarz and Anne R. Kroo and Anna J. Miller and Michel J. F. Digonnet and David A. B. Miller and Olav Solgaard},
      year={2025},
      eprint={2506.19242},
      archivePrefix={arXiv},
      primaryClass={physics.optics},
      url={https://arxiv.org/abs/2506.19242}, 
}

@article{Pai2019,
	author = {Pai, Sunil and Bartlett, Ben and Solgaard, Olav and Miller, David A. B.},
	doi = {10.1103/PhysRevApplied.11.064044},
	issue = {6},
	journal = {Phys. Rev. Appl.},
	month = {Jun},
	numpages = {18},
	pages = {064044},
	publisher = {American Physical Society},
	title = {Matrix Optimization on Universal Unitary Photonic Devices},
	url = {https://link.aps.org/doi/10.1103/PhysRevApplied.11.064044},
	volume = {11},
	year = {2019}}

@article{Wilkinson2021,
	author = {T. A. Wilkinson and C. E. Maurer and Colin J. Flood and Gerry H. Lander and Steve Chafin and E. B. Flagg},
	journal = {The Review of scientific instruments},
	pages = {093101},
	title = {Complete Stokes vector analysis with a compact, portable rotating waveplate polarimeter.},
	url = {https://api.semanticscholar.org/CorpusID:233231652},
	volume = {92 9},
	year = {2021}}

@article{Savory2010,
	author = {Savory, Seb J.},
	doi = {10.1109/JSTQE.2010.2044751},
	journal = {IEEE Journal of Selected Topics in Quantum Electronics},
	number = {5},
	pages = {1164-1179},
	title = {Digital Coherent Optical Receivers: Algorithms and Subsystems},
	volume = {16},
	year = {2010}
}

@book{Kaminow2010,
  editor    = {Kaminow, Ivan P. and Li, Tingye and Willner, Alan E.},
  title     = {Optical Fiber Telecommunications VB: Systems and Networks},
  edition   = {5th},
  year      = {2010},
  publisher = {Academic Press, an imprint of Elsevier},
  address   = {Amsterdam ; Boston},
  isbn      = {978-0-12-374172-1}
}

@article{Gisin2002,
	author = {Gisin, Nicolas and Ribordy, Gr\'egoire and Tittel, Wolfgang and Zbinden, Hugo},
	doi = {10.1103/RevModPhys.74.145},
	issue = {1},
	journal = {Rev. Mod. Phys.},
	month = {Mar},
	numpages = {0},
	pages = {145--195},
	publisher = {American Physical Society},
	title = {Quantum cryptography},
	url = {https://link.aps.org/doi/10.1103/RevModPhys.74.145},
	volume = {74},
	year = {2002}}

@book{Goldstein2011,
  author = {Goldstein, Dennis H.},
  edition = {3},
  publisher = {CRC Press},
  title = {Polarized Light},
  year = {2011},
  doi = {10.1201/b10436}
}

@article{Pusenkova2025,
	author = {Pusenkova, Anastasiia and Bagramyan, Aram and Larochelle, Patrick and Galstian, Van Vladimir and Galstian, Tigran},
	date = {2025/07/05},
	doi = {10.1038/s41598-025-07344-4},
	id = {Pusenkova2025},
	isbn = {2045-2322},
	journal = {Scientific Reports},
	number = {1},
	pages = {23996},
	title = {Polarimetric imaging with high spatial resolution},
	url = {https://doi.org/10.1038/s41598-025-07344-4},
	volume = {15},
	year = {2025}}

@article{Falk2014,
	author = {T{\"o}ppel, Falk and Aiello, Andrea and Marquardt, Christoph and Giacobino, Elisabeth and Leuchs, Gerd},
	date = {2014/07/16},
	doi = {10.1088/1367-2630/16/7/073019},
	isbn = {1367-2630; },
	journal = {New Journal of Physics},
	number = {7},
	pages = {073019},
	publisher = {IOP Publishing},
	title = {Classical entanglement in polarization metrology},
	url = {https://doi.org/10.1088/1367-2630/16/7/073019},
	volume = {16},
	year = {2014}}

@article{Fu2014,
	author = {Yunfei Fu and Tong Ye and Weijie Tang and Tao Chu},
	doi = {10.1364/PRJ.2.000A41},
	journal = {Photon. Res.},
	month = {Jun},
	number = {3},
	pages = {A41--A44},
	publisher = {Optica Publishing Group},
	title = {Efficient adiabatic silicon-on-insulator waveguide taper},
	url = {https://opg.optica.org/prj/abstract.cfm?URI=prj-2-3-A41},
	volume = {2},
	year = {2014}}

@article{Valdez2024,
	author = {Carson G. Valdez and Sunil Pai and Payton Broaddus and Olav Solgaard},
	doi = {10.1364/OL.517492},
	journal = {Opt. Lett.},
	month = {May},
	number = {9},
	pages = {2373--2376},
	publisher = {Optica Publishing Group},
	title = {High-efficiency vertically emitting coupler facilitated by three wave interaction gratings},
	url = {https://opg.optica.org/ol/abstract.cfm?URI=ol-49-9-2373},
	volume = {49},
	year = {2024}}

@inproceedings{Valdez2023,
	author = {Carson G. Valdez and Sunil Pai and Payton Broaddus and Olav Solgaard},
	booktitle = {CLEO 2023},
	journal = {CLEO 2023},
	pages = {JW2A.66},
	publisher = {Optica Publishing Group},
	title = {Triple-Etch Grating for Near Perfect Coupling at Normal Incidence},
	url = {https://opg.optica.org/abstract.cfm?URI=CLEO_FS-2023-JW2A.66},
	year = {2023}}

@article{Melloni2024,
	author = {Martinez, Andres Ivan and Cavicchioli, Gabriele and Seyedinnavadeh, Seyedmohammad and Zanetto, Francesco and Sampietro, Marco and D'Acierno, Alessandro and Morichetti, Francesco and Melloni, Andrea},
	date = {2024/08/30},
	doi = {10.1038/s41598-024-70726-7},
	id = {Martinez2024},
	isbn = {2045-2322},
	journal = {Scientific Reports},
	number = {1},
	pages = {20178},
	title = {Self-adaptive integrated photonic receiver for turbulence compensation in free space optical links},
	url = {https://doi.org/10.1038/s41598-024-70726-7},
	volume = {14},
	year = {2024}}

@misc{Banzer2025,
      title={Passive Silicon Nitride On-Chip Polarimetry: Precise Polarization Detection with Imperfect Components}, 
      author={Christoph Stockinger and Natale G. Pruiti and Isaac Tribaldo and Jörg S. Eismann and Marc Sorel and Peter Banzer},
      year={2025},
      eprint={2512.03920},
      archivePrefix={arXiv},
      primaryClass={physics.optics},
      url={https://arxiv.org/abs/2512.03920}, 
}

@article{Butow22,
	author = {Johannes B\"{u}tow and J\"{o}rg S. Eismann and Maziyar Milanizadeh and Francesco Morichetti and Andrea Melloni and David A. B. Miller and Peter Banzer},
	doi = {10.1364/OPTICA.458727},
	journal = {Optica},
	month = {Aug},
	number = {8},
	pages = {939--946},
	publisher = {Optica Publishing Group},
	title = {Spatially resolving amplitude and phase of light with a reconfigurable photonic integrated circuit},
	url = {https://opg.optica.org/optica/abstract.cfm?URI=optica-9-8-939},
	volume = {9},
	year = {2022}}

@article{Zaidi2024,
	author = {Zaidi, Aun and Rubin, Noah A. and Meretska, Maryna L. and Li, Lisa W. and Dorrah, Ahmed H. and Park, Joon-Suh and Capasso, Federico},
	date = {2024/07/01},
	doi = {10.1038/s41566-024-01426-x},
	id = {Zaidi2024},
	isbn = {1749-4893},
	journal = {Nature Photonics},
	number = {7},
	pages = {704--712},
	title = {Metasurface-enabled single-shot and complete Mueller matrix imaging},
	url = {https://doi.org/10.1038/s41566-024-01426-x},
	volume = {18},
	year = {2024}}

\end{document}